\documentclass[%
 aip,
 amsmath,amssymb,
 reprint,%
]{revtex4-1}

\usepackage{graphicx}
\usepackage{dcolumn}
\usepackage{bm}

\usepackage[utf8]{inputenc}
\usepackage[T1]{fontenc}
\usepackage{mathptmx}
\usepackage{etoolbox}

\makeatletter
\def\@email#1#2{%
 \endgroup
 \patchcmd{\titleblock@produce}
  {\frontmatter@RRAPformat}
  {\frontmatter@RRAPformat{\produce@RRAP{*#1\href{mailto:#2}{#2}}}\frontmatter@RRAPformat}
  {}{}
}%
\makeatother
\begin{document}

\preprint{AIP/123-QED}

\title{Connecting Q to TCF for MEMS and piezoelectric resonators}
\author{S. McHugh}%
 \email{smchugh@resonant.com.}
\affiliation{ 
Resonant - A Murata Company, San Mateo, California 94402 USA
}%

\date{\today}

\begin{abstract}
Two critical characteristics for any MEMS resonator are the quality factor ($Q$) and the temperature coefficient of frequency ($TCF$).  The connection between $Q$ and $TCF$ is demonstrated here with a phenomenological anharmonic oscillator model. Specifically, it is shown that the same nonlinear terms responsible for the $TCF$ set an upper limit on the resonator's $Q$.  A concise formula is found to estimate this loss and is shown to be closely related to Woodruff's formula for Akhiezer damping.  The use of this formula is illustrated by extending the model to an important class of MEMS; piezoelectric resonators.  Finally, the model is applied to published data for an AlN-on-diamond piezoelectric resonator. The focus here is on MEMS resonators, but the method should apply broadly to resonances with non-zero $TCF$.
\end{abstract}

\maketitle

\section{Introduction}
Akhiezer damping is understood to set the upper limit on the $Q$ of a MEMS resonator at high frequencies.\cite{Akhiezer, Atalaya, Rodriguez}  Qualitatively, it can be thought of as nonlinear loss mechanism that scatters energy from a primary to higher order modes.  Despite the limitations to its accuracy,\cite{Candler} Woodruff's formula\cite{Woodruff} concisely quantifies the upper limit on Q of an oscillating mode in a bulk material,
\begin{equation}
    \label{Wood}
    Q_W^{-1} = \frac{\gamma^2C_v\Omega\tau T}{\rho v^2}.
\end{equation}
$T$ is temperature, $\rho$ is the density, $v$ is speed of sound, $\gamma$ is the Gr\"uneisen parameter, $C_v$ is the heat capacity, $\Omega$ is the frequency of the excitation, and $\tau$ is thermal relaxation time.  With this detailed knowledge of the material properties one can calculate this fundamental loss for a material and estimate the upper limit of $Q$ for a given MEMS resonator under ideal circumstances.  However, MEMS resonators are often fabricated in highly strained states.  This happens most often as an accident of fabrication from remnant compressive stress resulting in bowed or buckled membranes.\cite{bachtold}  It can also occur from deliberate strain engineering, as is the case for energy harvesting applications \cite{MARIELLO2021105986,Kim, Yang} or dissipation dilution.\cite{Engelsen}  In each of these circumstances, the characterization of the Akhiezer damping can differ significantly from expectations based on non-strained calculations.

The nonlinear properties of the constituent MEMS materials also determine the thermal expansion, which in turn affects the resonant frequency.  Most materials expand as the temperature increases. In the case of a MEMS, BAW resonator, this leads to a decrease in resonant frequency, or a $TCF$ < 0.\cite{mi11070630} In principle, the $TCF$ of a BAW resonator can be predicted with detailed knowledge of the nonlinear material properties of each material in the resonator. In practice, this is rarely done.  Rather, the experimentally determined thermal expansion and temperature dependence of the materials are used explicitly.\cite{Smith, Ha}  Although this approach is often practical, it obscures the nonlinear origin of the temperature dependence.  

In this paper, the microscopic origins of loss are ignored and a phenomenological model is introduced to connect a resonator's $TCF$ and the upper limit on its $Q$.  Using a classical, anharmonic oscillator with one degree of freedom as a model for a MEMS resonator, it is shown that the same nonlinear terms required to model its $TCF$ also set an upper limit on its $Q$. In particular, the upper limit on $Q$ is shown to be roughly inversely proportional to $TCF$.  This formula for $Q$ is also shown to be closely related to Woodruff's formula (eq. \ref{Wood}) for Akhiezer damping.

This anharmonic oscillator model should be broadly applicable, but the focus here is on MEMS resonators.  In particular, it is extended to the widely used class of MEMS; piezoelectric resonators.  The essential step is mapping the coupled acoustic (mechanical) and electric degrees of freedom to a single anharmonic oscillator.  The results from published data for an AlN-on-diamond piezoelectric resonator\cite{Yamamoto} are analyzed with this model and found to be in reasonable agreement.  

There are shortcomings.  Since the model is empirical, and the parameters are not tied to directly to physical quantities that can be measured by other methods, its predictive power is limited.  For example, once a measured resonator is characterized with this model, it will not always be clear how to engineer a resonator with different characteristics, e.g., lower $|TCF|$ or higher $Q$.  However, the model should help clarify the underlying physics of the resonator and reveal whether or not such improvements are possible.  

\section{anharmonic Oscillator}
\label{AOsection}
First, consider a classical, driven, damped, harmonic oscillator with the familiar equation of motion
\begin{equation}
    \label{HO}
    m\ddot x = -kx -\frac{m\omega_0}{Q}\dot x + j,
\end{equation}
where $x$ is the displacement from equilibrium, $m$ is the mass, $k_0$ is the spring constant, $\omega_0 = \sqrt{k/m}$, $j$ is the driving force, and the damping  force is proportional to $Q^{-1}$.  Here, and throughout this paper, the parameters describing the oscillator are assumed to be the bare values for $T=0$.  Temperature effects will be modeled explicitly rather than with temperature-dependent, effective parameters.  

Contact with a thermal heat reservoir with temperature $T$ results in a random, white driving force on the oscillator.  Assuming $\Delta$ is such a force, eq. \ref{HO} can be written in frequency space as
\begin{equation}
    \label{SHOw}
    x(\omega) = \frac{\Delta/m}{\omega_0^2-\omega^2+i\frac{\omega_0\omega}{Q}}.
\end{equation}
The equipartition theorem requires $\left<x^2\right> = \frac{k_BT}{m\omega_0^2}$, where the angle brackets indicate average over long times, and $k_B$ is the Boltzmann constant.\cite{Pathria}  According to eq. \ref{SHOw}, this becomes
\begin{equation}
\label{x12}
    \left<x^2\right> = \int d\omega\left<\frac{\Delta^2/m^2}{(\omega_0^2-\omega^2)^2 + \frac{\omega_0^2\omega^2}{Q^2}}\right> = \frac{k_BT}{m\omega_0^2}.
\end{equation}
Since $\Delta$ is assumed to be white, the integrand is sharply peaked at $\omega=\omega_0$.  This implies that thermal excitation results in oscillations primarily at $\omega_0$.  Assume now that the driving force is the sum of the random thermal, a static $(\omega=0)$, and a force with frequency $\omega$, i.e., $j = j_0 + j_1\mbox{cos}\omega t + \Delta$.  In this case, $\left<x^2\right>$ is the sum of each source squared,
\begin{equation}
    \label{x2}
    \left<x^2\right> = \left(\frac{j_0}{m\omega_0^2}\right)^2 + \left(\frac{j_1Q}{m\omega_0\omega} \right)^2 + \frac{k_BT}{m\omega_0^2}.
\end{equation}
Although this discussion may sound obvious, it is helpful to write eq. \ref{x2} explicitly for the formulation of $Q$ to follow.

\subsection{Thermal expansion}
Without a static driving force, the average value of $x$ for a harmonic oscillator is zero, $\left<x\right> = 0$, i.e., there is no thermal expansion. All MEMS oscillators suffer the effects of temperature.  In order to model these effects, nonlinear terms must be added to eq. \ref{HO}.  Anticipating the requirements at room temperature, second and third order nonlinearities are added,
\begin{equation}
    \label{AO}
    m\ddot x = -kx + bx^2 + cx^3 -\frac{m\omega_0}{Q}\dot x + j,
\end{equation}
where $b$ and $c$ and have units $\left[\frac{kg}{ms^2}\right]$ and $\left[\frac{kg}{m^2s^2}\right]$, respectively. In the appendix, it is shown that including the nonlinear parameters gives the temperature dependence
\begin{equation}
    \label{AO_expansion}
    \left<x\right> = \frac{b}{k^2}k_BT\left(1 + \frac{3c}{k^2}k_BT\right),
\end{equation}
which should be treated as the static displacement from equilibrium, i.e., thermal expansion of the resonator.

\subsection{Temperature dependence of resonant frequency}
The nonlinear terms also lead to a resonance frequency shift and the generation of harmonics of the driving force.  To illustrate, assume $T=0$ and consider $j = Re\left(j_0+j_1e^{i\omega t}\right)$.  Searching for perturbative solutions, insert $j$ and $x = Re\left(x_0 + x_1e^{i\omega t} + x_2e^{i2\omega t}+\ldots\right)$, where $x_0\gg x_1\gg x_2\ldots$, into eq. \ref{AO} and group by powers of $e^{i\omega t}$.  

The static displacement is calculated first.  Assuming weak nonlinearities, i.e., $k \gg bx_0+cx_0^2,$ 
\begin{equation}
    x_0 \approx \frac{j_0}{k}.
\end{equation}
In the absence of an explicit static driving force, eq. \ref{AO_expansion} implies there may be a temperature-dependent, effective driving force, $j_0 = \frac{b}{k}k_BT\left(1+
3\frac{c}{k^2}k_BT\right).$

Next, consider the amplitude of the first harmonic given by 
\begin{equation}
    \label{1stHarmonic}
    x_1 = Re\left(\frac{j_1/m}{\omega_0^2 -2\frac{b}{m}x_0 -3\frac{c}{m}x_0^2+ i\frac{\omega_0\omega}{Q} - \omega^2}\right),
\end{equation}
which clearly has a resonance at 
\begin{equation}
    \label{1stHarmonicResonance}
    \omega^2 = \omega_0^2 -2\frac{b}{m}x_0 -3\frac{c}{m}x_0^2.
\end{equation}

$T\neq0$ causes a static bias on the resonant frequency giving it a dependence on temperature.  Specifically, $x_0$ of  eq. \ref{1stHarmonicResonance} can be replaced with $\left<x\right>$ from eq. \ref{AO_expansion} giving,
\begin{eqnarray}
    \label{omega T}
    \omega^2 &\approx& \omega_0^2 - 2\frac{b^2}{mk^2}k_BT -9\frac{b^2c}{mk^4}k_B^2T^2\nonumber\\
    &=& \omega_0^2 -2\frac{\bar b^2}{\omega_0^4}k_BT -9\frac{\bar b^2\bar c}{\omega_0^8}k_B^2T^2
\end{eqnarray}
where terms of order $T^3$ and higher are ignored.  Also, the scaled versions of $b$ and $c$ are introduced to make bookkeeping easier
\begin{equation}
    \bar b = \frac{b}{m^{3/2}},
\end{equation} and 
\begin{equation}
    \bar c = \frac{c}{m^2}.
\end{equation}  
The three parameters $\omega_0^2$, $\bar b^2$, and $\bar c$ can be deduced from measurements of $\omega^2$ over a sufficiently broad range of temperatures.

\subsection{Temperature-dependent, nonlinear loss}
Finally, consider the term proportional to $e^{i2\omega t}$ with amplitude
\begin{equation}
\label{SecondHarmonic}
    x_2 = Re\left(\frac{\frac{b}{m}x_1^2 + 3\frac{c}{m}x_0x_1^2}{\omega_0^2-2\frac{b}{m}x_0-3\frac{c}{m}x_0^2 - 2i\frac{\omega_0\omega}{Q} -4\omega^2} \right).
\end{equation}
Driving the oscillator at $\omega$, generates a displacement oscillating at $2\omega$.  The energy stored in the second harmonic (and higher harmonics) can be treated as the energy lost from the primary mode at $\omega.$  To estimate the upper limit imposed on $Q$ due to this nonlinear loss, the energy in each mode must be estimated.  

In a steady state, the energy of the oscillator described by eq. \ref{AO} is given 
\begin{equation}
    \label{E}
    E = \frac{m}{2}\dot x^2 + \frac{m\omega_0^2}{2}x^2 - \frac{b}{3}x^3 - \frac{c}{4}x^4.
\end{equation}
The steady state energy of the first harmonic is given approximately by
\begin{equation}
    E_1 = \frac{1}{2}m\omega^2x_1^2.
\end{equation}
The steady state energy of the second harmonic is
\begin{eqnarray}
    E_2 &=& \frac{1}{2}m\omega^2x_2^2\left<4\mbox{sin}^2(2\omega t) + \mbox{cos}^2(2\omega t) \right>\nonumber \\
    &\approx& \frac{5}{4}m\omega^2x_2^2,
\end{eqnarray}
where $\left<\right>$ indicates time averaging.

Using the definition of $Q^{-1} = \frac{1}{2\pi}\frac{\left<energy\mbox{ }lost\right>}{\left<energy\mbox{ }stored\right>}$, the upper limit on $Q$ imposed by nonlinearities is given by
\begin{equation}
    \label{Q}
    Q^{-1} = \frac{2}{5}\frac{x_2^2}{x_1^2}.  
\end{equation}
Of course, one could add the energy of the 3rd order and higher modes contributing to $\left<energy\mbox{ }lost\right>$ and estimate an even lower, upper bound on Q.  Assume also there are no other sources contributing to loss and Q is determined entirely by nonlinear loss.

Using eq. \ref{1stHarmonicResonance} for $\omega^2$ in eq. \ref{SecondHarmonic} gives
\begin{eqnarray}
\label{squared x2}
        x_2^2 &=& \frac{\left(\frac{b}{m}x_1^2 + 3\frac{c}{m}x_0x_1^2\right)^2}{9\omega^4+4\frac{\omega_0^2\omega^2}{Q^2}}\nonumber \\
        &\approx& \frac{1}{9\omega^4}\left(\frac{b}{m}x_1^2 + 3\frac{c}{m}x_0x_1^2\right)^2.
\end{eqnarray}
The approximation above comes from assuming $Q \gg 1$.  Without this assumption, the following would become an \emph{implicit} function of $Q$.  Using eq. \ref{squared x2} for $x_2^2$ in eq. \ref{Q} gives
\begin{equation}
    \label{QA1}
    Q^{-1} = \frac{2}{45}\frac{x_1^2}{\omega^4}\left(\frac{b}{m}+3\frac{c}{m}x_0\right)^2.
\end{equation}
This loss is explicitly nonlinear with the dependence on the mode amplitudes $x_0$ and $x_1$.  However, even in the case where the driving forces are small, a non-zero temperature will generate loss from the nonlinear terms. Assume the driving force is small, i.e., $j_1^2\ll \frac{m\omega_0^2}{Q}k_BT$.  Using eq. \ref{x12} for $x_1^2$, and eq. \ref{AO_expansion} for $x_0$, the expression for $Q_A^{-1}$ becomes
\begin{equation}
    \label{preInvQ}
    Q_A^{-1} = \frac{2}{45}\frac{k_BT}{\omega^4k}\left[\frac{b}{m}+3\frac{cb}{mk^2}k_BT\left(1 + 3\frac{c}{k^2}k_BT\right) \right]^2.
\end{equation}
Eq. \ref{omega T} is used to write $\omega^{-4}$ as
\begin{equation}
    \omega^{-4} = \omega_0^{-4}\left[1+4\frac{b^2}{k^3}k_BT+\ldots\right].
\end{equation}
Inserting this into eq. \ref{preInvQ} and dropping terms of higher order than $T^2$ gives
\begin{equation}
    \label{invQ}
    Q_A^{-1} = \frac{2}{45}\frac{\bar b^2}{\omega_0^6}k_BT\left[1 + \left(4\frac{\bar b^2}{\omega_0^6} +6\frac{\bar c}{\omega_0^4}\right)k_BT\right]
\end{equation}
$Q_A$ is meant to indicate the upper limit analogous to that set by Akhiezer damping.  Again, three parameters in this expression, $\omega_0, \bar b^2,$ and $\bar c$, can be completely determined by a measurement of $\omega$ vs. temperature, using eq. \ref{omega T}.  An example is provided below in Sec. \ref{ComparisonSection}.

\subsection{Comparison with Woodruff's formula for Akhiezer damping}
At low temperatures, where only the first term of eq. \ref{invQ} is relevant, it is nearly identical to Woodruff's formula for $Q_W^{-1}$ (eq. \ref{Wood}).  To make this more clear, first re-write $Q_W^{-1}$ in an alternative form using the coefficient of thermal expansion, $\alpha_v$ and the specific heat capacity, $C_p$,
\begin{equation}
    \label{Wood_rewrite}
    Q_W^{-1} = \frac{\alpha_v^2v^2}{C_p}\left(\Omega\tau\right)T.
\end{equation}

Assume $\left<E\right> \approx k_BT$ for a weakly nonlinear oscillator.  In this case, $C_p = \frac{1}{m}\frac{d\left<E\right>}{dT} = \frac{k_B}{m}$.  Using eq. \ref{AO_expansion} for $\left<x\right>$ at low temperatures, note that
\begin{equation}
    \omega_0^2\left(\frac{d\left<x\right>}{dT}\right)^2 = \omega_0^2\left(\frac{bk_B}{k^2} \right)^2
\end{equation}
can be used in place of $\alpha_v^2v^2$ in eq. \ref{Wood_rewrite}.  This gives Woodruff's formula for a single anharmonic oscillator as
\begin{equation}
    \label{Wood_rewrite2}
    Q_W^{-1} = \left(\Omega\tau\right)\frac{b^2}{k^3}k_BT.
\end{equation}

Compare this to eq. \ref{invQ} in the low temperature limit
\begin{equation}
    \label{invQ_linear}
    Q_A^{-1} = \frac{2}{45}\frac{b^2}{k^3}k_BT.
\end{equation}  
Eqs. \ref{Wood_rewrite2} and \ref{invQ_linear} differ only by their prefactors.  In this sense, eq. \ref{invQ} may be treated as the phenomenological equivalent to Woodruff's formula for Akhiezer damping extended to higher temperatures.

\subsection{$Q_A$ and $TCF$ at low temperatures}
It is common for MEMS researchers to report $\omega$ as a function temperature difference with respect to room temperature (e.g., ref. \onlinecite{Colombo}),
\begin{equation}
    \label{Omega}
    \omega = \Omega_0 + \bar{A}(T-T_{room}) + \bar{B}(T-T_{room})^2 +\ldots,
\end{equation}
where $\Omega_0$ is the resonant frequency at $T_{room}$, and $\bar{A}, \bar{B}, \ldots$ are constants.  The room temperature, linear $TCF$ is then defined as
\begin{equation}
    TCF_{room} = \frac{1}{\Omega_0}\frac{d\omega}{dT} = \frac{\bar{A}}{\Omega_0}.
\end{equation}  
In the low temperature limit, which is of interest here, it is more convenient to define the frequency dependence on absolute temperature as
\begin{equation}
\label{omega}
\omega = \omega_0 + AT + BT^2 +\ldots,
\end{equation}
where $\omega_0$ is again the resonant frequency at $T=0$.  In this case, the zero temperature, linear $TCF$ is defined as 
\begin{equation}
    TCF = \frac{1}{\omega_0}\frac{d\omega}{dT} = \frac{A}{\omega_0}.
\end{equation}
Squaring eq. \ref{omega} and comparing it to eq. \ref{omega T} in the low temperature limit reveals 
\begin{equation}
    -2\frac{\bar b^2k_B}{\omega_0^4} = 2\omega_0^2\left(\frac{1}{\omega_0}\frac{d\omega}{dT} \right).
\end{equation}
Substituting this into eq. \ref{invQ_linear} gives that particularly simple form 
\begin{equation}
    Q_A^{-1} \approx  \frac{2}{45}|TCF|T.
\end{equation}
In other words, a large $|TCF|$ leads to a low $Q_A$ for the resonator.  

Note, in the low temperature limit, $Q_A$ will generally not have a simple dependence on the coefficients defined with respect to room temperature as in eq. \ref{Omega}.

 \section{\label{sec:level1}Piezoelectric resonators} 
 The connection between $Q_A$ and $TCF$ described above applies to any anharmonic oscillator described by eq. \ref{AO}.  In this section, a piezoelectric resonator is modeled as such an anharmonic oscillator. 
 
 \begin{figure}
\includegraphics[width=0.4\textwidth]{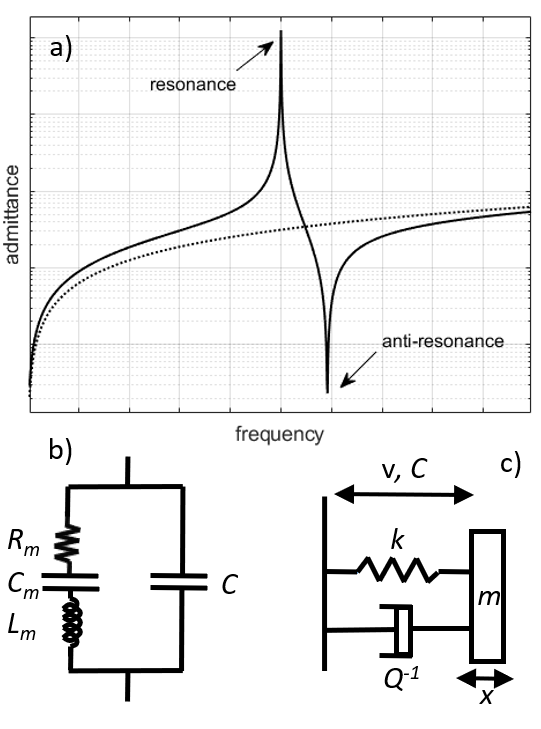}
\caption{\label{fig:bvd} a) Typical electrical admittance of a piezoelectric resonator (solid) compared to a simple capacitor (dotted).  b) Butterworth-van Dyke equivalent circuit for piezoelectric resonator. c) Schematic for coupled, electro-mechanical model. }
\end{figure}

The admittance (impedance$^{-1}$) for a generic piezoelectric resonator are shown in Figure \ref{fig:bvd}a.  The essential, small-signal characteristics of this resonator are captured by the Butterworth-van Dyke (BvD) model \cite{Butterworth, vanDyke} represented by the circuit drawn in fig. \ref{fig:bvd}b.  This BvD model can be derived from treating the piezoelectric resonator as a simple harmonic oscillator coupled to a capacitor governed by the coupled differential equations\cite{McHugh1, McHugh2}
\begin{eqnarray}
    \label{SHO}
    m\ddot{x} = -k x -\frac{\sqrt{mk}}{Q}\dot{x} + \alpha v \\
    \label{Ohm's}
    C v = q - \alpha x
\end{eqnarray}
The two degrees of freedom are the displacement, $x$ and voltage, $v$.  The mechanical or acoustic parameters are mass, $m$, spring constant, $k$, and resonator quality factor, $Q$.  $C$ is the capacitance and $q$ is the external charge.  Electrical resistance is treated as an external source of loss.  It is immaterial to the results presented here and is ignored.  The two differential equations are coupled through $\alpha$, the piezoelectric coupling. This is shown schematically in fig. \ref{fig:bvd}c.  

Eqs. \ref{SHO} and \ref{Ohm's} can be reduced to a single equation for the displacement of a harmonic oscillator
\begin{equation}
\label{reducedSHO}
    \ddot{x} = -\tilde\omega_0^2x - \frac{\omega_0}{Q}\dot{x} +\frac{\alpha}{mC}q
\end{equation}
where $\tilde\omega^2 = \omega_0^2+\frac{\alpha^2}{mC}$ and $\omega_0^2 = \frac{k}{m}$.
Fourier transformed solutions to \ref{reducedSHO} can be inserted into \ref{Ohm's} to give
\begin{equation}
    \label{Cap}
    C\left[\frac{\tilde\omega^2 + i\frac{\omega_0}{Q}\omega-\omega^2}{\omega_0^2 + i\frac{\omega_0}{Q}\omega-\omega^2}\right] v = q
\end{equation}
Evidently, a piezoelectric resonator may be thought of as a capacitor with a resonance frequency, $\omega_0$ and anti-resonance frequency, $\tilde\omega_0$.  The admittance for this resonator is then given by the BvD model where, $C_m=\alpha^2/k$, $L_m=m/\alpha^2$, $R_m = \frac{\sqrt{mk}}{\alpha^2Q}$, while $C$ is the same for both.  The model is over-specified with one more parameter than is required to match the data.  $m$ is treated as the free parameter and the critical results below will not depend on its value. 

For most common engineering purposes, all of these parameters are deduced by fitting the BvD model admittance to measurements of piezoelectric resonators measured near room temperature.  However, the resonator parameters, $m, k, C,$and $ \alpha$ discussed here will be treated as the low temperature, $T=0$ values.  

\subsection{Nonlinear corrections}
For large amplitude displacements and voltages, nonlinear corrections to eqs. \ref{SHO} and \ref{Ohm's} must be included. (See ref. \onlinecite{daniel} for an example of measured nonlinear parameters for a piezoelectric resonator.)  Since the linear version can be reduced to a single degree of freedom, $x$, $x$-dependent corrections to $k, C,$ and $\alpha$ are introduced as
\begin{eqnarray}
    \label{SHO NL}
    m\ddot{x} &=& -k(1-k_1x-k_2x^2) x - \frac{m\omega_0}{Q}\dot{x}\nonumber \\
    &+& \alpha(1-\alpha_1x-\alpha_2x^2) v \\
    \label{Ohm's NL}
    C(1-C_1x-C_2x^2)v &=& q - \alpha(1-\alpha_1x -\alpha_2x^2) x.
\end{eqnarray}
Since $m$ is assumed to be a free parameter, no correction is necessary. Each of the second order terms, $k_1, C_1$, and $\alpha_1$ have dimensions $m^{-1}$.  The third order terms $k_2, C_2$, and $\alpha_2$ have dimensions $m^{-2}$.  

Omitting terms of order $x^4$ and higher, eq. \ref{Ohm's NL} can be written as
\begin{eqnarray}
    \label{Ohm's NL approx}
    C_0v \approx Q +\left(qC_1-\alpha\right)x + \left(qC_2+QC_1^2-\alpha C_1+\alpha\alpha_1 \right)x^2\nonumber \\
    +\left(2qC_1C_2-\alpha C_2-\alpha C_1^2+\alpha\alpha_1C_1+\alpha\alpha_2 \right)x^3.
\end{eqnarray}
Inserting this into eq. \ref{SHO NL} and again omitting terms $x^4$ and higher gives
\begin{equation}
    \label{NLO}
    m\ddot{x} \approx -ax + bx^2 +cx^3 -\frac{\omega_0}{Q}\dot{x} + \frac{\alpha}{C}q, 
\end{equation}
where 
\begin{eqnarray}
a &=& k + \frac{\alpha^2}{C}+\frac{\alpha}{C}q\left(\alpha_1-C_1 \right)\\
b &=& k_1k + \frac{\alpha^2}{C}\left(2\alpha_1-C_1\right)\nonumber \\
&+& \frac{\alpha}{C}q\left(C_2+C_1^2-\alpha_1C_1-\alpha_2\right)\\
c &=& k_2k+\frac{\alpha^2}{C}\left(2\alpha_1C_1+2\alpha_2-C_2-C_1^2-\alpha_1^2\right)\nonumber \\
&+& \frac{\alpha}{C}q\left(2C_1C_2-\alpha_1C_2-\alpha_1C_1^2-\alpha_2C_1\right).
\end{eqnarray}
Eq. \ref{NLO} is the sought-after version of eq. \ref{AO} for a piezoelectric resonator.  The results of the perturbative approach used in Section \ref{AOsection} can now be applied directly to a piezoelectric resonator.  There are two differences worth noting.  One, the apparent resonance frequency of this oscillator actually corresponds to the \textit{anti}-resonance, $\tilde\omega$ of the piezoelectric resonator. (See fig. \ref{fig:bvd}.)  Two, this apparent anti-resonance frequency is explicitly dependent on the charge, or driving force, $\frac{\alpha}{C}q$.  For the previous results for anharmonic oscillators to hold for piezoelectric resonators, the external charge must be small.

Solutions to eq. \ref{NLO} describe the gross acoustic dynamics of piezoelectric resonator.  Once known, they can then be inserted into eq. \ref{Ohm's NL approx} to model the corresponding electrical behavior.

\subsection{Comparison to measurements}
\label{ComparisonSection}
Yamamoto et al.\cite{Yamamoto} performed detailed measurements of a piezoelectric, MEMS resonator from 300 to 5K.  They report the resonance $Q$ vs. $T$ for a AlN-on-diamond, SAW device with resonance frequency near 5 GHz.  These are convenient data to illustrate the use and validity of eq. \ref{invQ}.

\begin{figure}
\includegraphics[width=0.45\textwidth]{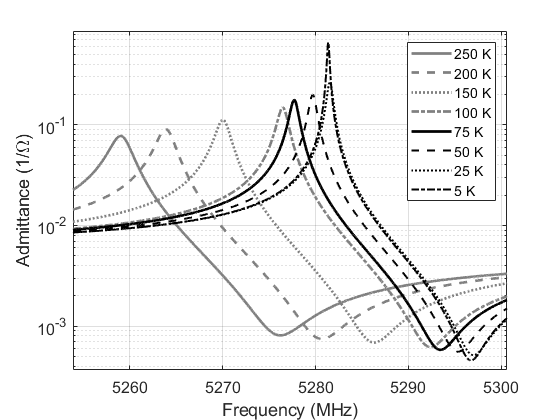}
\caption{\label{admittance data} Admittance reconstructed from BvD parameters reported in ref. \onlinecite{Yamamoto} for a piezoelectric resonator at eight temperatures.  The resonance and anti-resonance frequencies, along with the associated Qs increase as $T$ decreases.}
\end{figure}

\begin{figure}
\includegraphics[width=0.45\textwidth]{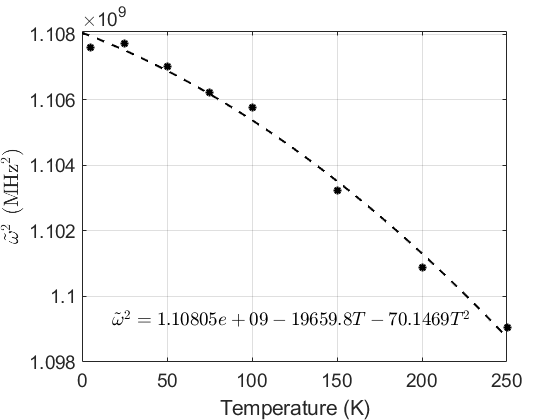}
\caption{\label{w vs T data} Square of the anti-resonance  vs. temperature from fig. \ref{admittance data}.  The dashed line is a quadratic, least-squares fit to the data obeying the inset equation.}
\end{figure}

\begin{figure}
\includegraphics[width=0.45\textwidth]{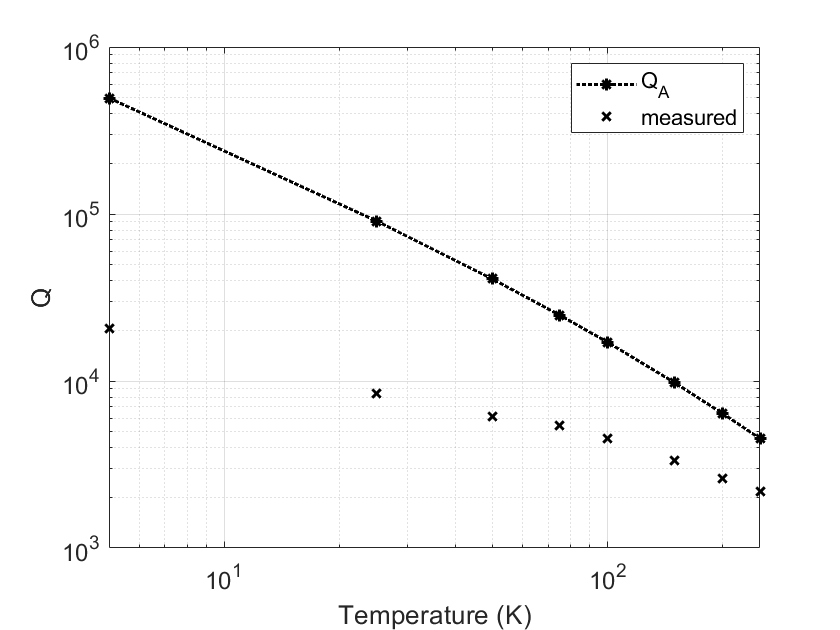}
\caption{\label{Q vs T data} Resonance Q vs. temperature. The measured resonance $Q$ reported in ref. \onlinecite{Yamamoto} is show with x markers.  The dotted line shows the upper limit on resonance $Q_A$ determined from the frequency vs. temperature data. }
\end{figure}

The anti-resonance frequency, $\tilde\omega$ is required for input to eq. \ref{invQ}, but was not reported explicitly in ref. \onlinecite{Yamamoto}.  The authors did, however, report the BvD parameters for the SAW resonator taken at each temperature.  The electrical admittance of the resonator at each temperature can be reconstructed using the reported BvD parameters.  Fig. \ref{admittance data} shows the reconstructed responses.\cite{anomalous}  The anti-resonance frequencies are deduced from these admittance curves.  The square of the anti-resonance frequencies are shown in fig. \ref{w vs T data}  as a function of $T$.  Also shown is a quadratic, least-squares fit.  $\tilde\omega_0^2$, $\bar b^2$, and $\bar c$ are determined from the coefficients of this equation.  These serve as inputs to eq. \ref{invQ}, giving the estimation for $Q_A^{-1}$.  $R_m$ is deduced from $Q_A$ as, $R_m = L_m\omega_0Q_A^{-1}$ and inserted back into the BvD models replacing the resistor values reported in ref. \onlinecite{Yamamoto}.  As the final step, the admittance is reconstructed for each temperature and the resonance $Q$ tabulated.  These are shown with the dotted line of fig. \ref{Q vs T data}.  Also shown are the measured resonance $Q$.

$Q_A$ is more than an order of magnitude larger than measured $Q$ at 5 K.  Evidently, the thermal, nonlinear loss is not a limiting factor at low temperatures for this resonator.  However, near room temperature, the measured $Q$ quite close to $Q_A$.  Note, the precise value of $Q_A$ is sensitive to the fitting parameters (shown in fig. \ref{w vs T data}).  As such, more accurate $\tilde\omega^2$ vs. $T$ could produce a significantly different estimation for $Q_A$.  

\section{Discussion and conclusion}
If the upper bound on $Q$ is orders of magnitude larger than measured values, it informs the researcher to look elsewhere for $Q$-limiting effects.  In situations where it is close to the measured values, it should be more useful.  In such cases, nonlinear conditions, e.g., nonlinear material properties, large static displacements, and geometric nonlinearities, must be limiting the loss and also affecting the $TCF$.  

An obvious limitation of this model is the requirement that resonator obey eq. \ref{omega T} from the temperature of interest down to low temperatures.  Often, piezoelectric resonators used for applications requiring high-stability are engineered to have nearly zero $TCF$ at room temperature.\cite{zeroTCF}  Such a resonator would most likely obey eq. \ref{omega T} only at substantially lower temperatures, where the estimation for $Q_A$ should apply. 

Finally, it is worth restating, this model is phenomenological and involves no microscopic details of the resonator.  As such, it is expected to apply broadly to resonances with a temperature dependent frequency. For instance, resonant circuits with a nonlinear capacitor\cite{cmut1, cleland},  resonant circuits with a nonlinear inductor\cite{SC1, SC2}, and even the magnetic resonance frequencies of NV centers in diamond\cite{NV1, NV2} all have resonance frequencies that vary with temperature.  Regardless of the microscopic origin, the nonlinearities causing the temperature dependent frequency shift should impose an upper limit on $Q$ for these diverse resonators. 

\begin{acknowledgments}
I thank P. J. Turner and A .N. Cleland for helpful discussions.
\end{acknowledgments}

\appendix

\section{Temperature dependence of $\left<x\right>$}
\label{appendix}
The method described in ref. \onlinecite{Feynman} is followed here to  deduce the temperature dependence of $\left<x\right>$ for the oscillator described by eq. \ref{AO}.

Ignoring the dissipation and source terms, the equations of motion (eq. \ref{AO}) can be derived from eq. \ref{E} treating $E$ as the Hamiltonian, $H$.  Non-zero $T$ will cause a displacement $\left<x\right> = A$.  Denote the Helmholtz free energy for eq. \ref{E} as $F$.  $A$ can be deduced by substituting $x = y+A$ and minimizing $F$ with respect to $A$.  However, the nonlinear terms in $H$ make calculating the expectation values difficult.  Instead, the calculations can be simplified by taking advantage of the theorem\cite{Feynman}
\begin{equation}
    F \leq F_0 + \left<H - H_0\right>_0,
\end{equation}
where, $H_0$ is any conveniently chosen Hamiltonian, $\left<\right>_0$ refers to the average taken with respect to $H_0$, and $F_0$ is the free energy of $H_0$.  

Let,
\begin{equation}
    H_0 = \frac{p^2}{2m} + \frac{1}{2}k(x-A)^2.
\end{equation}
$A$ is the displacement of the harmonic oscillator to be determine, i.e., $A = \left<x\right>.$  Now, evaluate
\begin{eqnarray}
    \left<H-H_0\right>_0 &=& \left<\frac{k}{2}x^2 - \frac{b}{3}x^3 - \frac{c}{4}x^4 -\frac{k}{2}(x-A)^2\right>_0 \nonumber\\
    &=& \left<\frac{k}{2}A^2 - Aby^2 -\frac{3}{2}cA^2y^2 - \frac{c}{4}y^4 \right>_0,
\end{eqnarray}
where $y = x-A$.  Also, $\left<x^n\right>_0 = 0$ for odd values of $n$, and values of $A^3$ and higher order are ignored.  

Now, $F_0 + \left<H - H_0\right>_0$ may be minimized with respect to $A$ to find that $A$ must satisfy
\begin{eqnarray}
    A &=& \frac{b\left<y^2\right>_0}{k - 3c\left<y^2\right>_0}\\
    &\approx& \frac{b}{k}\left<y^2\right>_0\left(1+3\frac{c}{k}\left<y^2\right>_0\right)
\end{eqnarray}
where $\frac{\partial F_0}{\partial A}=0$ and $3c\left<y^2\right>_0\ll k.$  Since $\left<y^2\right>_0 = k_BT/k$, $A$ can be written as
\begin{eqnarray}
        A &=& \frac{b}{k^2}k_BT\left(1 + \frac{3c}{k^2}k_BT\right)\\
        &=& \frac{b}{m^2\omega_0^4}k_BT\left(1 + \frac{3c}{m^2\omega_0^4}k_BT\right)
\end{eqnarray}

\nocite{*}
\bibliography{aipsamp}

\begin{thebibliography}{30}
\bibitem [{ano()}]{anomalous}%
  \BibitemOpen
  \href@noop {} {}\bibinfo {note} {The admittance reconstructed from the $T=300$ K BvD parameters appear anomalous and are omitted from the analysis here.}\BibitemShut {Stop}%
\end{thebibliography}

\end{document}